\documentclass[final,5p,times,twocolumn]{elsarticle} 
\DeclareMathAlphabet{\mathpzc}{OT1}{pzc}{m}{it}
\usepackage{lineno}
\usepackage{setspace}
\usepackage{graphicx}
\usepackage{amssymb}
\usepackage{amsmath}
\usepackage{amsthm}
\usepackage[colorlinks=true]{hyperref}
\usepackage[all]{hypcap}
\usepackage{mathrsfs}

\usepackage[percent]{overpic}
\usepackage{fixltx2e}
\usepackage{epsfig}
\usepackage{verbatim}
\usepackage{multirow}
\usepackage{float}
\usepackage{array}

\newcommand{\D}{\mathrm{d}}
\newcommand{\tot}{\mathrm{tot}}
\newcommand{\tof}{\mathrm{ToF}}
\newcommand{\dtc}{\mathrm{det}}
\newcommand{\dps}{\mathrm{dep}}
\newcommand{\sml}{\mathrm{sim}}
\newcommand{\mn}{\mathrm{min}}
\newcommand{\mx}{\mathrm{max}}
\newcommand{\el}{\mathrm{el}}
\newcommand{\E}{\mathcal{E}}
\newcommand{\prim}{\mathrm{prim}}
\newcommand{\scat}{\mathrm{scat}}
\newcommand{\cs}{\,\xi\,}
\newcommand{\A}{\mathcal{A}}
\newcommand{\cm}{\mathrm{cm}}
\newcommand{\lab}{\mathrm{lab}}
\newcommand{\cpt}{\mathrm{cap}}

\journal{Nuclear Instruments and Methods A}
\begin{document}
\begin{frontmatter}

\title{An improved method for estimating the neutron background in measurements of neutron capture reactions}


\author[a]{P.~\v{Z}ugec\corref{cor1}}\ead{pzugec@phy.hr}
\author[a]{D.~Bosnar}
\author[b]{N.~Colonna}
\author[c]{F.~Gunsing}

\address[a]{Department of Physics, Faculty of Science, University of Zagreb, Croatia}
\address[b]{Istituto Nazionale di Fisica Nucleare, Sezione di Bari, Italy}
\address[c]{Commissariat \`{a} l'\'{E}nergie Atomique (CEA) Saclay - IRFU, Gif-sur-Yvette, France}

\author{The n\_TOF Collaboration\fnref{fn1}}
\cortext[cor1]{Corresponding author. Tel.: +385 1 4605552}
\fntext[fn1]{www.cern.ch/ntof}

\begin{abstract}
The relation between the neutron background in neutron capture measurements and the neutron sensitivity related to the experimental setup is examined. It is pointed out that a proper estimate of the neutron background may only be obtained by means of dedicated simulations taking into account the full framework of the neutron-induced reactions and their complete temporal evolution. No other presently available method seems to provide reliable results, in particular under the capture resonances. An improved neutron background estimation technique is proposed, the main improvement regarding the treatment of the neutron sensitivity, taking into account the temporal evolution of the neutron-induced reactions. The technique is complemented by an advanced data analysis procedure based on relativistic kinematics of neutron scattering. The analysis procedure allows for the calculation of the neutron background in capture measurements, without requiring the time-consuming simulations to be adapted to each particular sample. A suggestion is made on how to improve the neutron background estimates if neutron background simulations are not available.
\end{abstract}

\end{frontmatter}

\section{Introduction}
\label{sec:chap1}

The background caused by scattering of neutrons off the irradiated sample is a serious issue in neutron capture experiments. Through subsequent neutron interactions with the materials surrounding the sample, secondary reaction products are created -- such as $\gamma$ rays and/or charged particles -- which may be detected alongside the capture $\gamma$ rays emitted from the sample, contributing to the total background. This particular contribution, referred to as the \emph{neutron background}, is most notable for the samples characterized by a large neutron scattering-to-capture cross section ratio. In general, neutron background is characteristic of environments which are strongly affected by the neutron scattering. It is intensified by the presence of any neutron-sensitive material in the immediate vicinity of the detectors, and especially by the detector proximity to the walls of the experimental hall. The neutron background is determined by two distinct components, one being the sample itself, serving as the primary neutron scatterer, and other being the sample-independent \emph{neutron sensitivity} related to the entire experimental setup. The neutron sensitivity may be generally defined as the detector response to reaction products created by the interaction of scattered neutrons with the surrounding materials. A nontrivial effect of the neutron sensitivity on the neutron background and, consequently, the entire capture measurement has been demonstrated by Koehler et al. \cite{koehler} in capture measurement on $^{88}$Sr, where the reduction of the neutron sensitivity of the experimental setup has led to significant improvements in the acquired capture data.


At the neutron time-of-flight facility n\_TOF at CERN, neutron sensitivity considerations have been followed since the start of its operation. This was reflected through the development of specially optimized C$_6$D$_6$ (deuterated benzene) liquid scintillation detectors, exhibiting a very low intrinsic neutron sensitivity \cite{plag}. However, the neutron background at the n\_TOF facility is heavily affected by the surrounding massive walls, serving as the prime candidates for the enhanced neutron scattering. Furthermore, the much higher neutron energies available from the n\_TOF spallation source introduce an additional contribution to the neutron background, when compared to the neutron sources based on electron LINACs, where the neutron energies are usually limited to $\sim$10~MeV. Details on the n\_TOF facility can be found in Refs.~\cite{carlos,ear2_1,ear2_2}.

Recently, GEANT4 \cite{geant4} simulations were developed for determining the neutron background in the measurements with C$_6$D$_6$ detectors at n\_TOF \cite{background}. The results of these simulations were first applied in the analysis of the experimental capture data for $^{58}$Ni \cite{ni58} and the analysis of the integral cross section measurement of the $^{12}$C($n,p$)$^{12}$B reaction \cite{carbon}. An earlier $^{58}$Ni capture measurement by Guber et al. \cite{guber} has already revealed that previous experimental results and adopted evaluations of the $^{58}$Ni capture cross section have been heavily affected by the neutron background, that was in the past inadequately suppressed or accounted for. At n\_TOF the neutron background was accurately determined by means of dedicated simulations benchmarked against the available measurements \cite{background}, and was subtracted from the $^{58}$Ni data \cite{ni58}.

The aim of this paper is to demonstrate that deriving the neutron background from the neutron sensitivity (Sections~\ref{sensitivity} and \ref{resonance}) or even the dedicated measurements (Section~\ref{natcarbon}) is not a trivial issue and requires a suitable procedure. We address the issue by developing an improved method for determining the neutron background, which is based on an advanced treatment of the simulated neutron sensitivity (Section~\ref{novel}). The improvements regard both the event tracking in the simulations and the subsequent data analysis. In particular, we propose to study the neutron sensitivity by keeping track of the total time delays between the neutron scattering off the sample and the detection of counts caused by the neutron-induced reactions. The limitations of the method are addressed in Section~\ref{uranium}. Section \ref{summary} summarizes the results and conclusions of this work. A detailed mathematical formalism underlying the proposed method is reported throughout the Appendices~\hyperlink{response_function}{A}, \hyperlink{analysis}{B}, and \hyperlink{formalism}{C}.

\section{Neutron sensitivity vs. the neutron background}
\label{sensitivity}

When comparing the neutron background to the neutron sensitivity, a clear distinction has to be made concerning the neutron energies. The \emph{primary neutron energy} is the true energy of the neutron (from the incident neutron beam) that has caused the reaction or the chain of reactions leading to the neutron background. The \emph{reconstructed energy} is the energy determined from the total time delay between the neutron production and the detection of secondary particles generated by the neutron-induced reactions. In case of the \emph{prompt counts}, caused by the reaction products immediately produced in the sample (e.g. $\gamma$ rays from neutron capture), the reconstructed energy is equal to the neutron kinetic energy, due to the total time delay being equal to the neutron time-of-flight. In case of neutron scattering inside the experimental hall or some other delay mechanism, such as the decay of radioactive products created by neutron-induced reactions, the total time delay may be large and may significantly affect the reconstructed neutron energy. For these \emph{delayed counts} contributing to the neutron background, the reconstructed energy will be lower than the primary neutron energy, often by orders of magnitude. While the reconstructed energy is experimentally accessible, the primary neutron energy is not, and can only be determined by simulations.

The neutron background estimation methods laid out in Sections~\ref{sensitivity}, \ref{resonance} and \ref{natcarbon} neglect the difference between the primary neutron energy $\E$ (before the scattering), the scattering energy $E_n$ (sampled in the simulations) and the reconstructed energy $E_\tof$. Hence, throughout these Sections the notation $E_n$ will be used as the universal one for the neutron energy. We follow this approach for consistency with Refs.~\cite{plag,background,frank}, freely combining the considerations strictly valid either for $\E$, $E_n$ or $E_\tof$. Starting from Section~\ref{novel}, these distinctions will be explicitly taken into account. In that, it should be noted that the neutron sensitivity has conventionally been expressed in terms of the scattering neutron energy \cite{plag,background,frank}. On the other hand, the neutron background -- as appearing in the experiments -- is a function of the reconstructed energy, suggesting at once an incompatibility between the two.

In order to calculate the neutron background from the neutron sensitivity, one needs to determine the neutron detection efficiency $\varepsilon_n$, i.e. the efficiency for detecting a neutron through the detection of particles produced in secondary neutron reactions. This is commonly achieved by running the dedicated simulations, wherein the neutrons are isotropically and isolethargically generated from a point source at the sample position. We note that the Pulse Height Weighting Technique \cite{pwht} has to be applied in calculating the efficiency, in order to compensate for the lack of correlations between $\gamma$ rays in the simulated \mbox{$\gamma$-ray} cascades following neutron captures. This issue has already been addressed in Ref.~\cite{frank}. Furthermore, the central role of applying the Pulse Height Weighting Technique to the simulated capture data was unambiguously confirmed in Ref.~\cite{background} by comparing the simulated and the experimental capture data for $^{197}$Au. A detailed description of the Pulse Height Weighting Technique applied at n\_TOF may be found in Ref.~\cite{pwht_ntof}.

We adopt the definition of the neutron sensitivity from \mbox{Refs. \cite{plag,background}}, which uses the ratio $\varepsilon_n/\varepsilon_\gamma^\mathrm{max}$, taking into account the maximum \mbox{$\gamma$-ray} detection efficiency $\varepsilon_\gamma^\mathrm{max}$ as an additional constant factor. In order to be able to use the weighted neutron detection efficiency $\varepsilon_n^{(w)}$, we further generalize the definition of the neutron sensitivity $S$, by introducing the average weighting factor $\langle w\rangle$:
\begin{linenomath}\begin{equation}
\label{eq2}
S(E_n)\equiv\frac{\varepsilon_n^{(w)}(E_n)}{\varepsilon_\gamma^{\mathrm{max}}\times\langle w\rangle}
\end{equation}\end{linenomath}
The weighted neutron detection efficiency $\varepsilon_n^{(w)}$ is:
\begin{linenomath}\begin{equation}
\varepsilon_n^{(w)}(E_n)=\frac{\sum_{i=1}^{\delta N_\dtc(E_n)}w_i(E_\dps)}{\delta N_\sml(E_n)}
\end{equation}\end{linenomath}
with $w_i$ as the appropriate weighting factors from the Pulse Height Weighting Technique, dependent on the energy $E_\dps$ deposited in detectors. $\delta N_\dtc(E_n)$ is the number of detected counts caused by neutrons of scattering neutron energy $E_n$, while $\delta N_\sml(E_n)$ is the total number of neutrons simulated at this energy. The average weighting factor $\langle w\rangle$ is obtained by taking into account all neutron energies sampled:
\begin{linenomath}\begin{equation}
\langle w\rangle=\frac{\sum_{E_n}\sum_{i=1}^{\delta N_\dtc(E_n)}w_i(E_\dps)}{\sum_{E_n}\delta N_\dtc(E_n)}
\end{equation}\end{linenomath}
It may be noted that without weighting ($w_i=1$ for all counts) the generalized neutron sensitivity reverts to the original $\varepsilon_n/\varepsilon_\gamma^\mathrm{max}$ ratio. The weighted efficiency $\varepsilon_n^{(w)}$ has been calculated for two C$_6$D$_6$ detectors used at n\_TOF. One is the modified version of a commercial Bicron detector and the other one was custom built at Forschungszentrum Karlsruhe and denoted as FZK detector \cite{plag}. For the sake of simplicity and clarity, in this paper we will only show the results for the Bicron detector, with the condition $E_\dps>200$~keV, as usually imposed on the experimental data. Furthermore, the reader's attention may be drawn to noticeable fluctuations apparent in multiple figures presented throughout this paper. With the exception of clearly recognizable resonances in the displayed spectra, the fluctuations are purely statistical in nature -- a simple consequence of a finite runtime dedicated to the computationally intensive simulations. They are also naturally enchanted by the application of the Pulse Height Weighting Technique and by a fine binning that was selected for displaying the data, in order to preserve the clear appearance of some of the very narrow resonances.

\begin{figure}[t!]
\includegraphics[width=1.\linewidth,keepaspectratio]{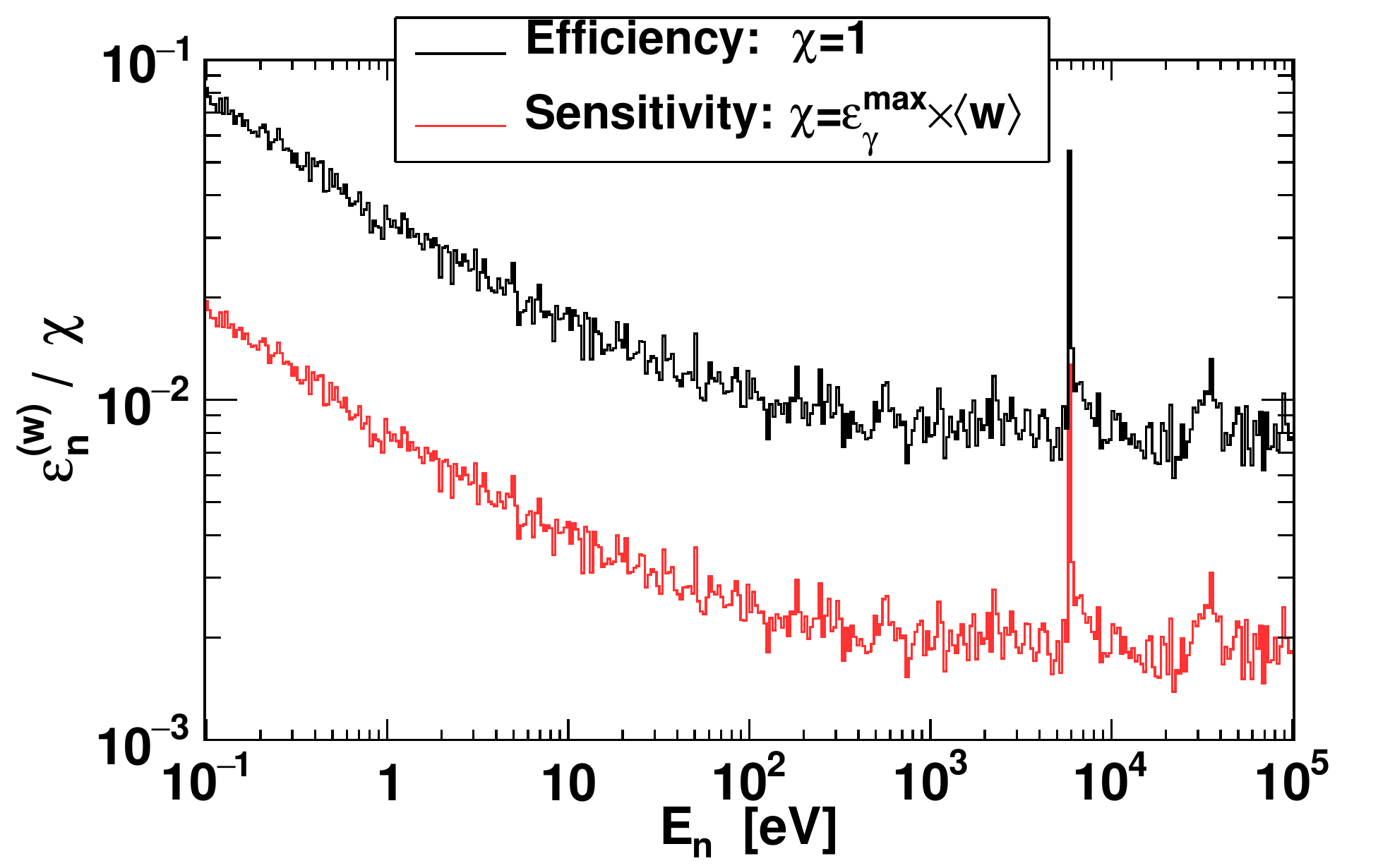}
\caption{(Color online) Weighted neutron detection efficiency of the modified Bicron detector used at n\_TOF, including the effect of the entire experimental hall. The corresponding generalized neutron sensitivity is also shown.}
\label{fig1}
\end{figure}

In accordance with the laid out considerations, Fig.~\ref{fig1} shows the weighted neutron detection efficiency of the Bicron detector. The generalized neutron sensitivity, offset by a constant factor $\varepsilon_\gamma^\mathrm{max}\times\langle w\rangle$, is also shown, since it will be used later on.

\begin{figure}[t!]
\includegraphics[width=1.\linewidth,keepaspectratio]{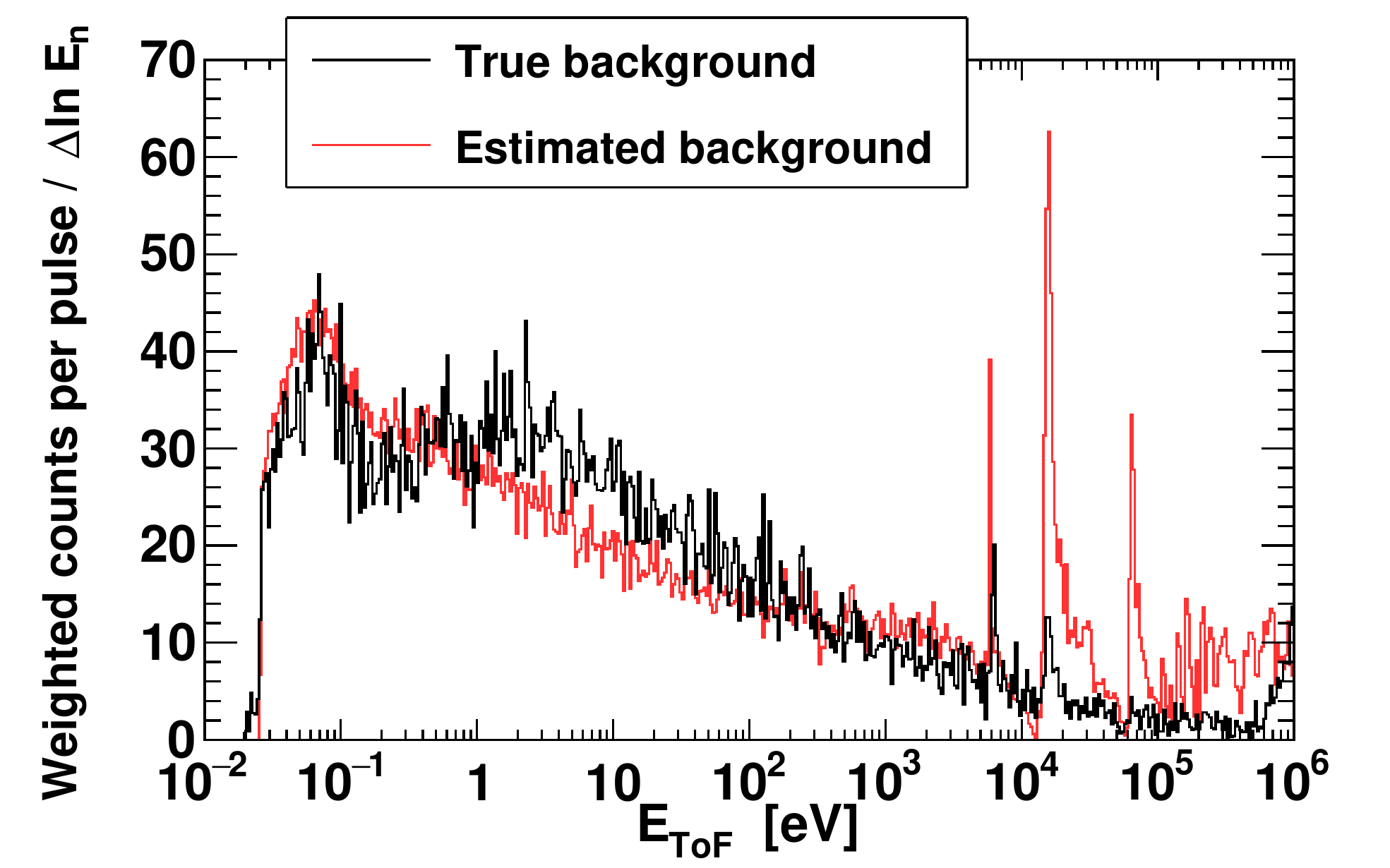}
\caption{(Color online) Comparison between the true weighted neutron background for $^{58}$Ni and the one estimated from the weighted neutron detection efficiency of the Bicron detector.}
\label{fig2}
\end{figure}

To estimate the neutron background in terms of the weighted counts per neutron bunch, the yield $Y_\mathrm{el}(E_n)$ of elastically scattered neutrons (i.e. the scattering probability) and the neutron flux $\phi(E_n)$ (normalized to the number of neutrons per neutron bunch) have to be taken into account. For purely illustrative purposes, we will assume a simple relation for the yield:
\begin{linenomath}\begin{equation}
\label{yield}
Y_\mathrm{el}(E_n)=\left(1-e^{-n\sigma_\mathrm{tot}(E_n)}\right)\frac{\sigma_\mathrm{el}(E_n)}{\sigma_\mathrm{tot}(E_n)}
\end{equation}\end{linenomath}
where $n$ is the areal density of the sample in number of atoms per unit surface. $\sigma_\mathrm{el}$ and $\sigma_\mathrm{tot}$ denote the elastic scattering cross section and the total cross section, respectively. This expression does not take into account the multiple scattering effects, the angular distribution of scattered neutrons nor the contribution from inelastic reactions, which may all be accounted for in dedicated simulations, as in Ref.~\cite{frank}. The weighted neutron background $B^{(w)}(E_n)$ may be expressed as:
\begin{linenomath}\begin{equation}
\label{estimate}
B^{(w)}(E_n)=\varepsilon_n^{(w)}(E_n)Y_\mathrm{el}(E_n)\phi(E_n)
\end{equation}\end{linenomath}
This estimate is compared in Fig. \ref{fig2} against the true neutron background for $^{58}$Ni, obtained from dedicated simulations \cite{background}. Outside the resonance region the scale of the true background is, indeed, very well reproduced by the estimated one. However, under the capture resonances the background calculated from the neutron detection efficiency is clearly overestimated. This is precisely because the estimated background is expressed in terms of the primary neutron energy instead of the reconstructed energy, thus missing the time delays following the neutron scattering off the sample. Since the background overestimation seems to be more pronounced for strong resonances, subtracting the neutron background estimated from the neutron detection efficiency may significantly affect the measured capture resonances. This is particularly troublesome because the strongest capture resonances constitute the dominant contribution to the Maxwellian averaged cross sections (MACS), which represent the basic input for astrophysical models of stellar nucleosynthesis.

\section{Neutron background under capture resonances}
\label{resonance}

As is shown in Fig.~\ref{fig2}, the neutron background estimated from the neutron sensitivity of the experimental setup exhibits strong resonances due to the elastic scattering cross section of a given sample. Since each capture resonance is accompanied by the corresponding resonant component in the elastic scattering cross section, an erroneous estimate of the neutron background may heavily affect the capture data. Here we discuss the correct method of estimating the neutron background, focusing on the illustrative example of the strong 15.3~keV resonance in the $^{58}$Ni($n,\gamma$) reaction.

The first estimate of the neutron background under the capture resonances relies on simple neutron sensitivity considerations. To demonstrate the limitations of this method, we benchmark it against the true neutron background under the 15.3 keV resonance (obtained by dedicated simulations; shown in Fig.~\ref{fig2}). We assume that the capture yield $Y_\cpt$ under the resonance is determined by its radiative width $\Gamma_\gamma$: $Y_\cpt\propto\Gamma_\gamma$. The reaction yield is translated into the number of the detected capture counts $C_\cpt$ through the average \mbox{$\gamma$-ray} detection efficiency $\varepsilon_\gamma$ as: $C_\cpt\propto\varepsilon_\gamma\Gamma_\gamma$. By the same reasoning, the yield of elastically scattered neutrons $Y_\el$ is determined by the neutron width $\Gamma_n$ as: $Y_\el\propto\Gamma_n$. The neutron background counts $C_\el$ are similarly affected by the normalized weighted neutron detection efficiency $\overline{\varepsilon}_n=\varepsilon_n^{(w)}/\langle w\rangle$, so that: $C_\el\propto\overline{\varepsilon}_n\Gamma_n$. In order to establish the link with the adopted definition of the generalized neutron sensitivity $S=\overline{\varepsilon}_n/\varepsilon_\gamma^\mathrm{max}$, we replace the average \mbox{$\gamma$-ray} detection efficiency $\varepsilon_\gamma$ by the maximum one $\varepsilon_\gamma^\mathrm{max}$. The relative contribution of the neutron background to the total number of counts measured at the given resonance may then be estimated as:
\begin{linenomath}\begin{equation}
\label{eq1}
\frac{C_\el}{C_\cpt+C_\el}\approx\frac{\overline{\varepsilon}_n\Gamma_n}{\varepsilon_\gamma^\mathrm{max}\Gamma_\gamma+\overline{\varepsilon}_n\Gamma_n}=\frac{S\Gamma_n}{\Gamma_\gamma+S\Gamma_n}
\end{equation}\end{linenomath}
For the $^{58}$Ni resonance at 15.3~keV the values from ENDF/B-VII.1 \cite{endf71} are: $\Gamma_\gamma$ = 1.104 eV and $\Gamma_n$ = 1354.062 eV. With the neutron sensitivity for the Bicron detector of $S=1.76\times10^{-3}$, determined by averaging the data from Fig.~\ref{fig1} within the resonance range from 14~keV to 17~keV, the ratio from Eq.~(\ref{eq1}) amounts to approximately 68\%. On the contrary, dedicated simulations of the neutron background indicate that when expressed in terms of the primary (note: \emph{not} reconstructed!) neutron energy, the neutron background between 14~keV and 17~keV amounts only to 24\% of the total detected counts. The difference relative to the result of Eq.~(\ref{eq1}) is due to the following reasons: (1)~in Eq.~(\ref{eq1}) the maximum \mbox{$\gamma$-ray} detection efficiency was used instead of the average one (which, in fact, lowers the estimated value); (2)~in small part because the neutron sensitivity was calculated assuming an isotropic distribution of scattered neutrons (isotropic in laboratory frame), instead of a more realistic one; (3)~most importantly, the elastic scattering cross section has a different shape than the neutron capture cross section, showing pronounced interference patterns and more extended resonance tails, strongly affecting not only the partial contribution to the yield of scattered neutrons within the limited energy range between 14~keV to 17~keV, but also the proportionality between the overall scattering yield and the neutron width, relative to the capture counterparts ($Y_\el/\Gamma_n\neq Y_\cpt/\Gamma_\gamma$). These simple considerations already indicate that a simplified approach, such as the one of Eq.~(\ref{eq1}), may lead to large errors in the estimates of the neutron background.

\begin{figure}[t!]
\includegraphics[width=1.\linewidth,keepaspectratio]{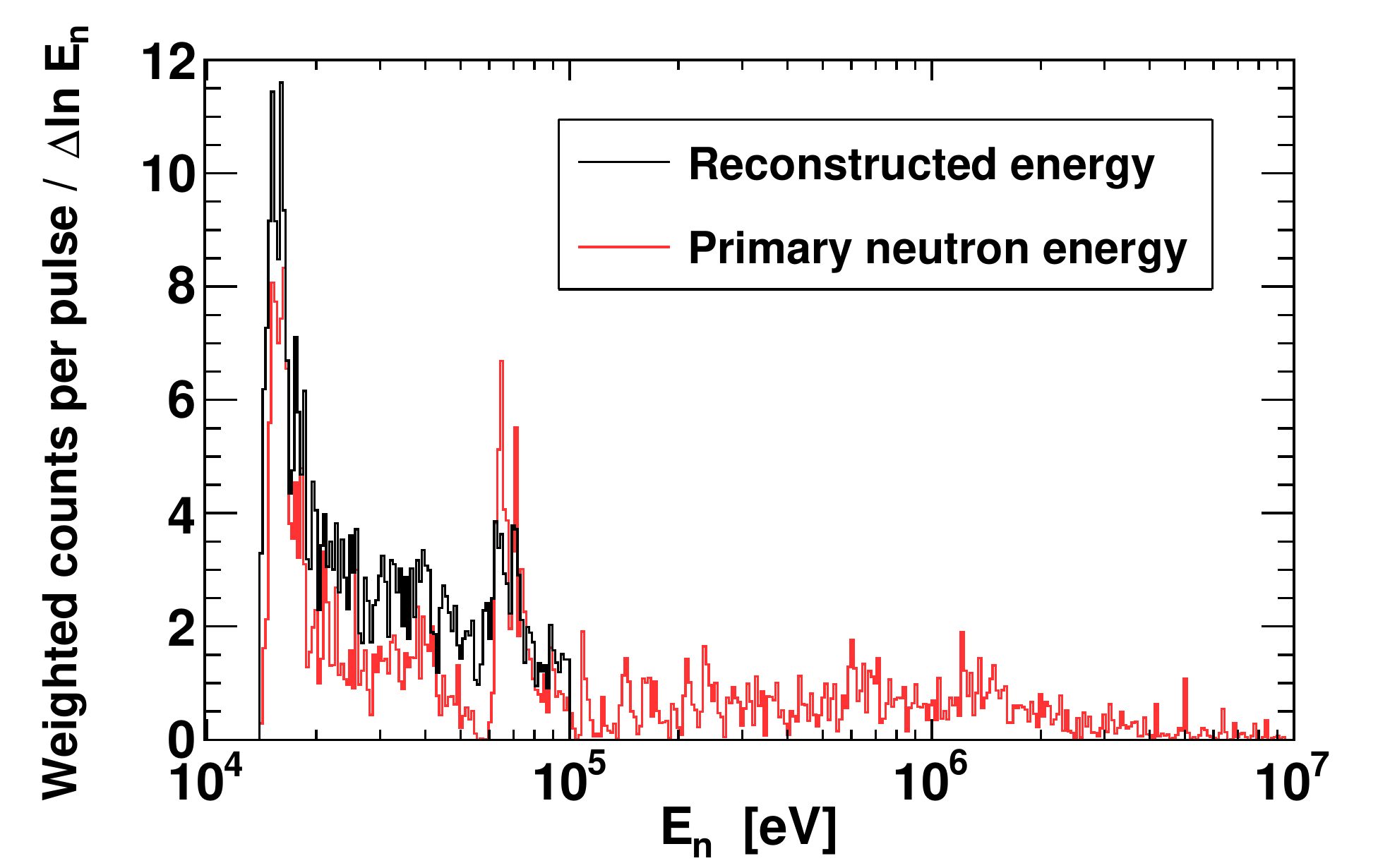}
\caption{(Color online) Weighted neutron background of $^{58}$Ni within the 14~keV -- 100~keV range, compared to the spectrum of primary energies for the neutrons that have caused it.}
\label{fig5}
\end{figure}

A very different value for the relative contribution of the neutron background is obtained if we consider the reconstructed neutron energy. The true neutron background, integrated between  14~keV and 17~keV, makes only 6.2\% of the total resonance area. Moreover, this reduced value consists of both the prompt and the delayed component, while the neutron sensitivity considerations apply only to the prompt component. (The delayed component was defined by considering events with more than 1\% relative difference between the primary neutron energy $\E$ and the reconstructed energy $E_\tof$, i.e. $E_\tof<0.99\,\E$. Increasing this relative difference to 10\% did not produce any notable changes between the two components.) If treated separately, the delayed component, which cannot be estimated on the basis of simple neutron sensitivity considerations, contributes 2.6\% to the 15.3~keV resonance, while the prompt component amounts to only 3.6\%, in contrast with the initial value of 24\% from the neutron background expressed as a function of the primary neutron energy.

\begin{figure}[t!]
\includegraphics[width=1.\linewidth,keepaspectratio]{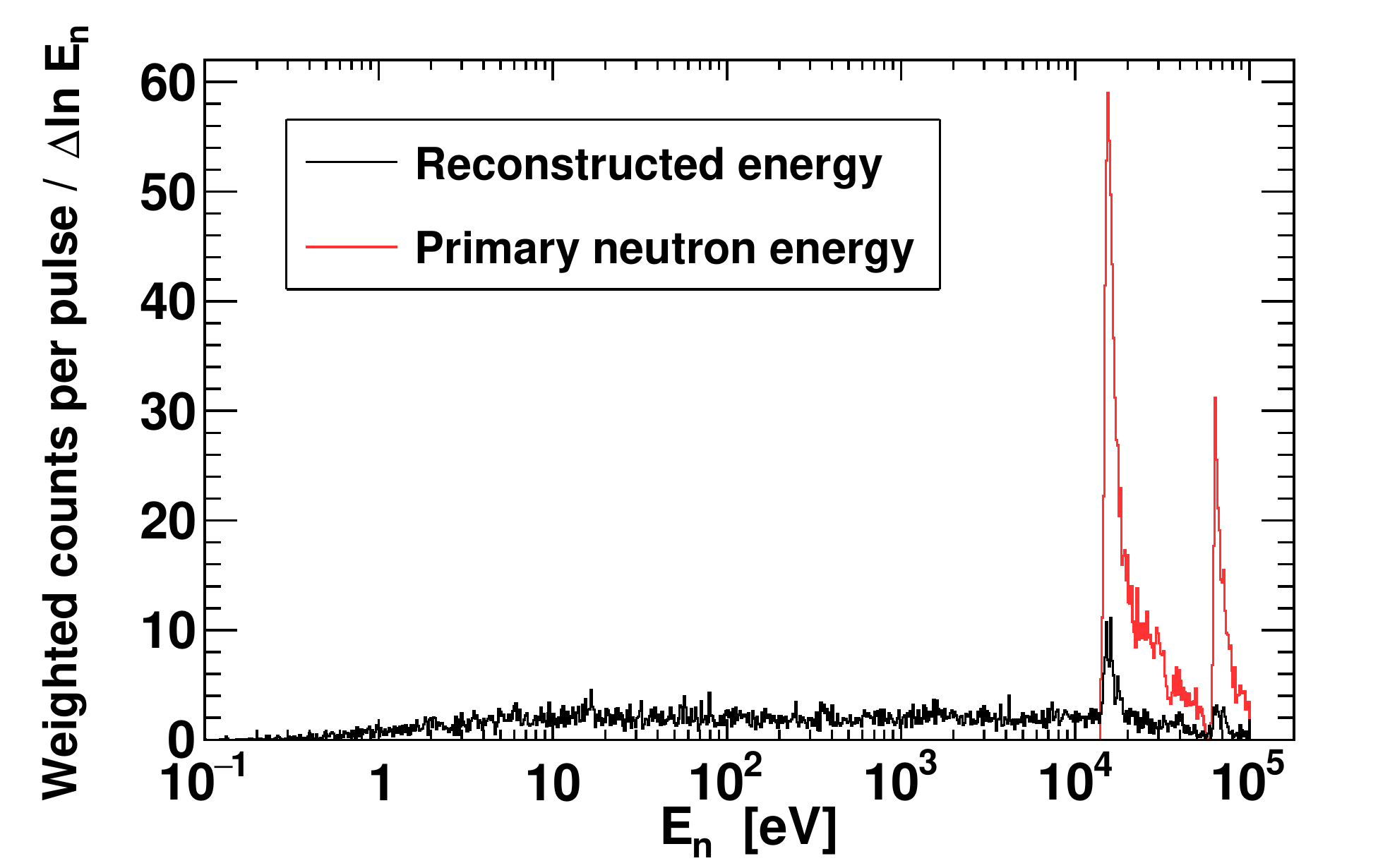}
\caption{(Color online) Weighted neutron background of $^{58}$Ni caused by neutrons with primary energies between 14~keV and 100~keV.}
\label{fig6}
\end{figure}

In order to understand the difference between the value of 24\% from the primary neutron energies and the value of 6.2\% from the reconstructed energies, we focus on the origin of the neutron background within the 14~keV -- 100~keV range, where two strong capture resonances (15.3~keV and 63.3~keV) are located. Figure \ref{fig5} compares the total neutron background in that region with the energy spectrum of primary neutrons that have generated it. Neutrons all the way up to 10~MeV contribute notably to the background below 100~keV, while the contribution from neutrons of higher energy is negligible.

On the other hand, Fig.~\ref{fig6} shows the background produced by neutrons of primary energies between 14~keV and 100~keV. The comparison of Figs.~\ref{fig5} and \ref{fig6} shows that a large fraction of the background in the 14~keV -- 100~keV range is produced by higher-energy neutrons. Simultaneously, the neutrons of primary energy from the considered range produce a background mostly contained at lower reconstructed energies, down to 0.1~eV. Only a small fraction of the background remains in the same energy range (in particular, only 15\%, as the ratio between previously quoted values of 3.6\% and 24\%). From both figures, together with Fig. \ref{fig2}, we reach the following conclusion: contrary to the past assumption, not even the prompt component of the neutron background under the capture resonances may be safely estimated from neutron sensitivity considerations alone. Instead, complete simulations of the neutron propagation throughout the experimental hall must be performed, taking the full temporal evolution of the neutron-induced reactions into account.


\section{Scaling the measured carbon background}
\label{natcarbon}

Another common method used for estimating the neutron background -- without relying on simulations, except for possible higher order corrections -- consists in scaling the neutron background measured with a neutron scatterer, such as a sample of natural carbon. This method is based on the assumption that the measured background is proportional to the yield of elastically scattered neutrons for a given sample: $B^{(w)}(E_n)\propto Y_\mathrm{el}(E_n)$, with the constant of proportionality assumed to be equal for both the scatterer and the sample under investigation. Under such assumptions the weighted background related to a given sample (we use the example background $B_{^{58}\mathrm{Ni}}^{(w)}(E_n)$ related to the $^{58}$Ni sample) may be estimated from the weighted background $B_{^\mathrm{nat}\mathrm{C}}^{(w)}(E_n)$, measured with the carbon sample, as:
\begin{linenomath}\begin{equation}
\label{carbon}
B_{^\mathrm{nat}\mathrm{C}}^{(w)}(E_n)=\frac{B_{^\mathrm{nat}\mathrm{C}}^{(w)}(E_n)}{Y_\mathrm{el}^{(^\mathrm{nat}\mathrm{C})}(E_n)}\times Y_\mathrm{el}^{(^{58}\mathrm{Ni})}(E_n)
\end{equation}\end{linenomath}
where $Y_\mathrm{el}^{(^{58}\mathrm{Ni})}(E_n)$ and $Y_\mathrm{el}^{(^\mathrm{nat}\mathrm{C})}(E_n)$ are the yields of elastically scattered neutrons for $^{58}$Ni and $^\mathrm{nat}$C, respectively.


\begin{figure}[t!]
\includegraphics[width=1.\linewidth,keepaspectratio]{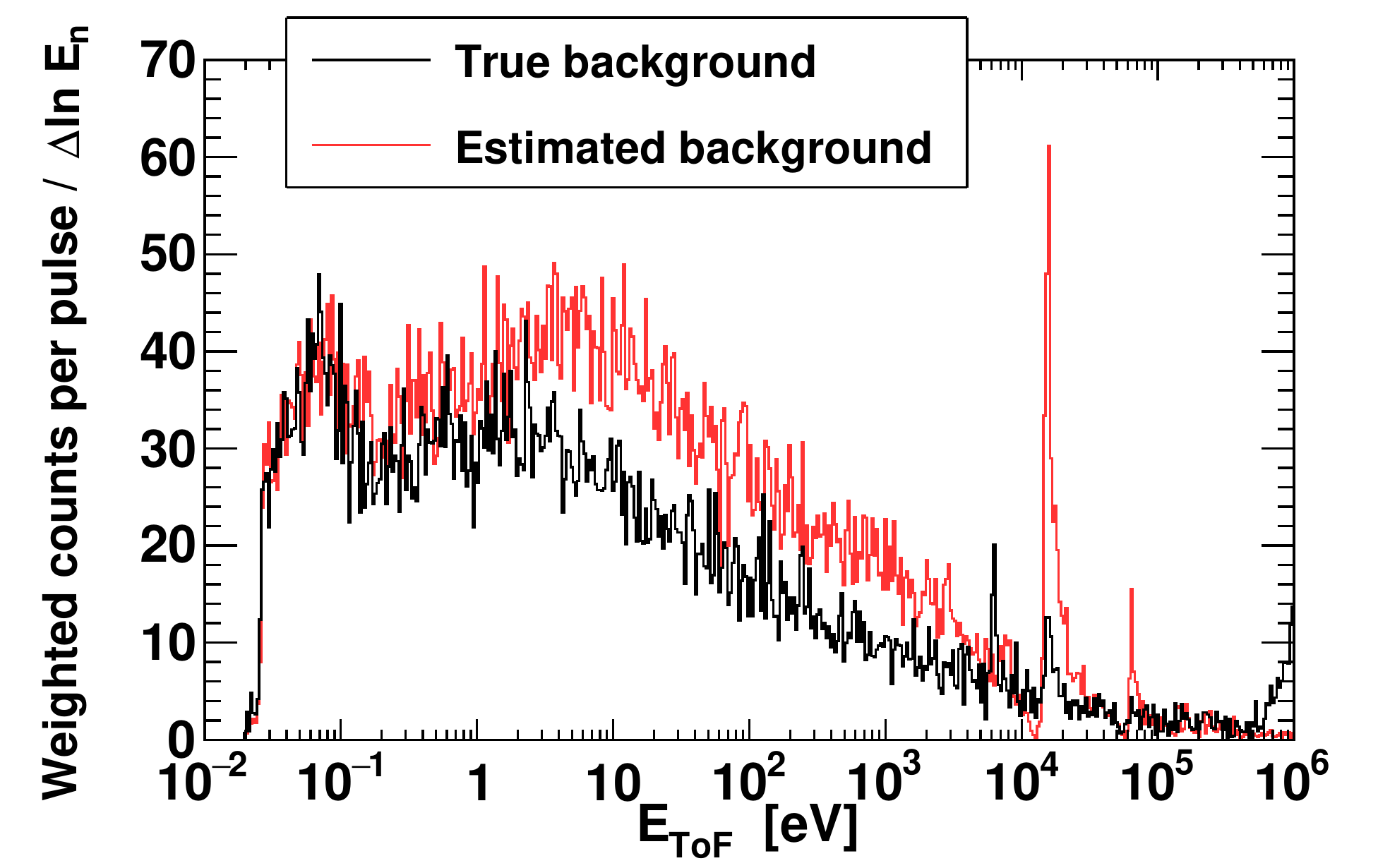}
\caption{(Color online) Comparison between the true weighted neutron background of $^{58}$Ni and the one estimated from the weighted neutron background of $^\mathrm{nat}$C sample.}
\label{fig8}
\end{figure}

Figure~\ref{fig8} compares the actual neutron background for $^{58}$Ni with the one estimated by Eq.~(\ref{carbon}), using the simulated neutron background for $^\mathrm{nat}$C. Under strong resonances the neutron background is again overestimated, for the same reasons that were covered in Section \ref{resonance}. However, a startling agreement may be observed outside the resonant region of $^{58}$Ni, which may be surprising at first, since the principle of scaling the backgrounds for different samples by the portion of elastically scattered neutrons (at a given primary neutron energy) can only be applied to the prompt components. Though, in principle, an excellent agreement outside the resonant region may not be \emph{a priori} expected, it may be understood from the fact that the sources of the neutron background other than the sample itself (i.e. the experimental components and the walls of the experimental area) are equal for both samples.

If Figs. \ref{fig2} and \ref{fig8} have shown anything, it is that in estimating the neutron background directly from the elastic scattering cross section, it is by far more appropriate to use the smoothed modification of the cross section, in which the (strong) scattering resonances have been flattened and replaced by the smooth sections connecting to the cross section at surrounding energies.

Unfortunately, Fig. \ref{fig8} may betray an overly optimistic picture, since for estimating the neutron background of $^{58}$Ni, the simulated neutron background of $^\mathrm{nat}$C was used, instead of the experimental one. The reason is that the measured carbon data from Experimental Area~1 of the n\_TOF facility reveal the pure neutron background only at the reconstructed energies above 1~keV \cite{background}. Below this energy the measurements are strongly affected by the detection of $\beta$ rays from the decay of radioactive $^{12}$B residuals produced by the inelastic $^{12}$C($n,p$)$^{12}$B reaction \cite{carbon}, opening above the reaction threshold of 13.6~MeV. This is an intrinsic feature of $^{12}$C itself and can not be projected to the other samples. For this reason, in Experimental Area~1 the neutron background for $^{58}$Ni (or any other sample) can not be estimated in the energy range below 1~keV directly from the measured carbon data. At the same time, any reliable extrapolation toward lower energies is extremely hard, especially since the shape of the background below 10~eV decidedly departs from the one set above 1~keV.

\section{Improved neutron background estimation procedure}
\label{novel}

The considerations up to this point reveal the limitations of the currently available methods for determining the neutron background. Herein, we propose an improved approach to estimating the neutron background, based on simulations of the complete experimental setup, barring the sample itself. These simulations need to be performed only once for a given experimental setup, with the output results -- reflecting the detector response to the scattered neutrons -- being adjusted to the scattering properties of a particular sample by means of a proper normalization technique. The simulations are basically identical to those used for the neutron sensitivity in Section~\ref{sensitivity} and are described in Ref.~\cite{background}. In short, the overall experimental setup is irradiated by neutrons isolethargically and isotropically emitted from a point source at the sample position. Note that the neutrons are emitted as if they have been already scattered off the sample. For this reason, we will treat them as the \emph{scattered neutrons}, as opposed to the \emph{primary neutrons} from the neutron beam. The distinction and its importance are elaborated in \mbox{\ref{considerations}}. The sampled neutron energies span the full energy range of the n\_TOF neutron beam: from thermal up to 10~GeV. Each count detected by any of the two C$_6$D$_6$ detectors is characterized by the following set of parameters extracted from simulations: $(E_n,E_\dps,T,\theta,\varphi)$. Here $E_n$ is the initial neutron energy, $E_\dps$ is the energy deposited in the detector, $T$ is the time delay between the neutron emission from the sample position and the detection of secondary particles, while $\theta$ and $\varphi$ are conventionally defined polar and azimuthal angles of the initial neutron emission, relative to the primary beam direction. The most important feature of the improved approach is that the neutron background is calculated from the time delays $T$, and expressed as a function of the reconstructed energy $E_\tof$, instead of the initially sampled energy $E_n$. In addition, a more involved data analysis will be included in the procedure.

\begin{figure}[t!]
\includegraphics[width=1.\linewidth,keepaspectratio]{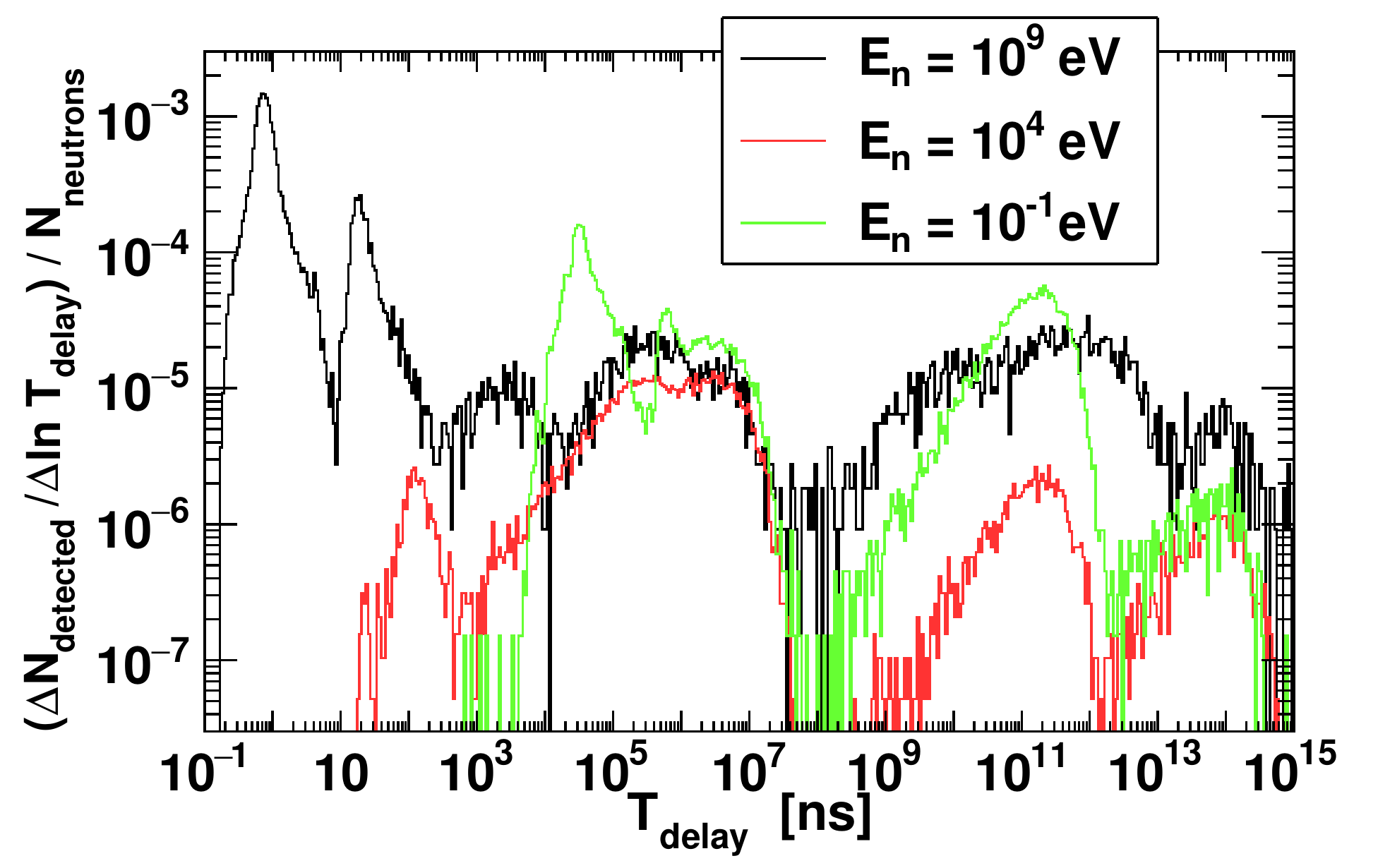}
\caption{(Color online) Detector response to neutrons isotropically scattered in the laboratory frame (as a function of time delay between the neutron scattering off the sample and the detection of secondary particles), for selected energies of scattered neutrons. See the main text for details.}
\label{fig11}
\end{figure}

Figure~\ref{fig11} shows examples of the detector response to neutrons isotropically scattered in the laboratory frame, for several selected scattering energies $E_n$, as functions of the time delay $T$. The detector response function $f_{E_n,\cs}^{(\scat)}(T)$ and its relation to the neutron background are explained in detail in \mbox{\ref{response_function}}. Fig.~\ref{fig11} shows the angle-integrated response $\int_{-1}^1 f_{E_n,\cs}^{(\scat)}(T)\,\D\cs$. The counts at very large time delays are caused by the detection of particles (mostly $\beta$ rays) emitted in the decay of long-lived radionuclides produced by neutron activation. Only counts with a time delay of up to $\sim$100~ms contribute significantly to the neutron background affecting the n\_TOF measurements (since $\sim$100~ms is the width of the data acquisition window at n\_TOF facility; the exact vale depends on the adopted sampling rate of the data acquisition system). The later counts may only contribute (with sharply decreasing probability) through the wrap-around process in subsequent neutron pulses.

The backbone of the new method is the proper normalization of the simulated data, which consists in weighting each detected count by the appropriate weighting factors $\mathcal{W}(*)$, dependent on more parameters than just the reconstructed energy $E_\tof$, which appears as the main argument of the neutron background. Since the cosine of the scattering angle plays a dominant role, in order to abbreviate relevant expressions we use the following notation:
\begin{linenomath}\begin{equation}
\label{short_cos}
\cs\equiv\cos\theta
\end{equation}\end{linenomath}
together with an abbreviation for the set of all relevant parameters:
\begin{linenomath}\begin{equation}
\label{star}
*\equiv \{E_\tof,E_n,E_\dps,\cs\}
\end{equation}\end{linenomath}
We report here the central expression for the weighting factors, while the complete mathematical formalism underlying the method is treated in detail in a series of Appendices (\hyperlink{response_function}{A}, \hyperlink{analysis}{B}, \hyperlink{formalism}{C}). In particular, an overview of considerations leading to the correct expression for the weighting factors $\mathcal{W}(*)$ is presented in \mbox{\ref{considerations}}, with the detailed derivation given in \mbox{\ref{derivation}}. Adopting the notation from Eqs.~(\ref{short_cos}) and (\ref{star}), the expression for the weighting factors may be written as:
\begin{linenomath}\begin{equation}
\label{weighting}
\mathcal{W}(*)=\frac{w(E_\dps)\, Y_\el(\E)\,\phi(\E)\,\eta_\E(\cs)\,\ln\frac{E_\mx}{E_\mn}}{N_\tot}\times\frac{E_n}{\E}\frac{\partial\E}{\partial E_n}
\end{equation}\end{linenomath}
Here $w(E_\dps)$ is the weighting function from the Pulse Height Weighting Technique. $Y_\el(\E)$ is the yield of elastically scattered neutrons, given by Eq.~(\ref{yield}), but dependent on the \emph{primary neutron energy} $\E$ from before the scattering off the sample. The primary neutron energy $\E$ is given as a function of the \emph{neutron scattering energy} $E_n$ and the scattering angle $\cs$ by Eq.~(\ref{primary}) in \mbox{\ref{considerations}}. The neutron flux $\phi(\E)$ is given in units of lethargy. The angular correction factor $\eta_\E(\cs)$ -- given by Eq.~(\ref{eq3}) in \mbox{\ref{considerations}} -- translates the isotropically simulated angular distribution $\A^{(\mathrm{iso})}(\cs)$ of the scattered neutrons into the more realistic laboratory distribution $\A_\E^{(\mathrm{lab})}(\cs)$:
\begin{linenomath}\begin{equation}
\label{ang}
\A_\E^{(\mathrm{lab})}(\cs)=\eta_\E(\cs)\times\A^{(\mathrm{iso})}(\cs)
\end{equation}\end{linenomath}
which is the relativistically transformed angular distribution of neutrons isotropically scattered in the neutron-nucleus center of mass frame. Finally, $N_\tot$ is the total number of simulated neutrons, with $E_\mn$ and $E_\mx$ as the minimum and maximum sampling energy, respectively. We remark that the derivative $\partial\E/\partial E_n$ may be easily calculated from Eq.~(\ref{primary}).

\begin{figure}[t!]
\includegraphics[width=1.\linewidth,keepaspectratio]{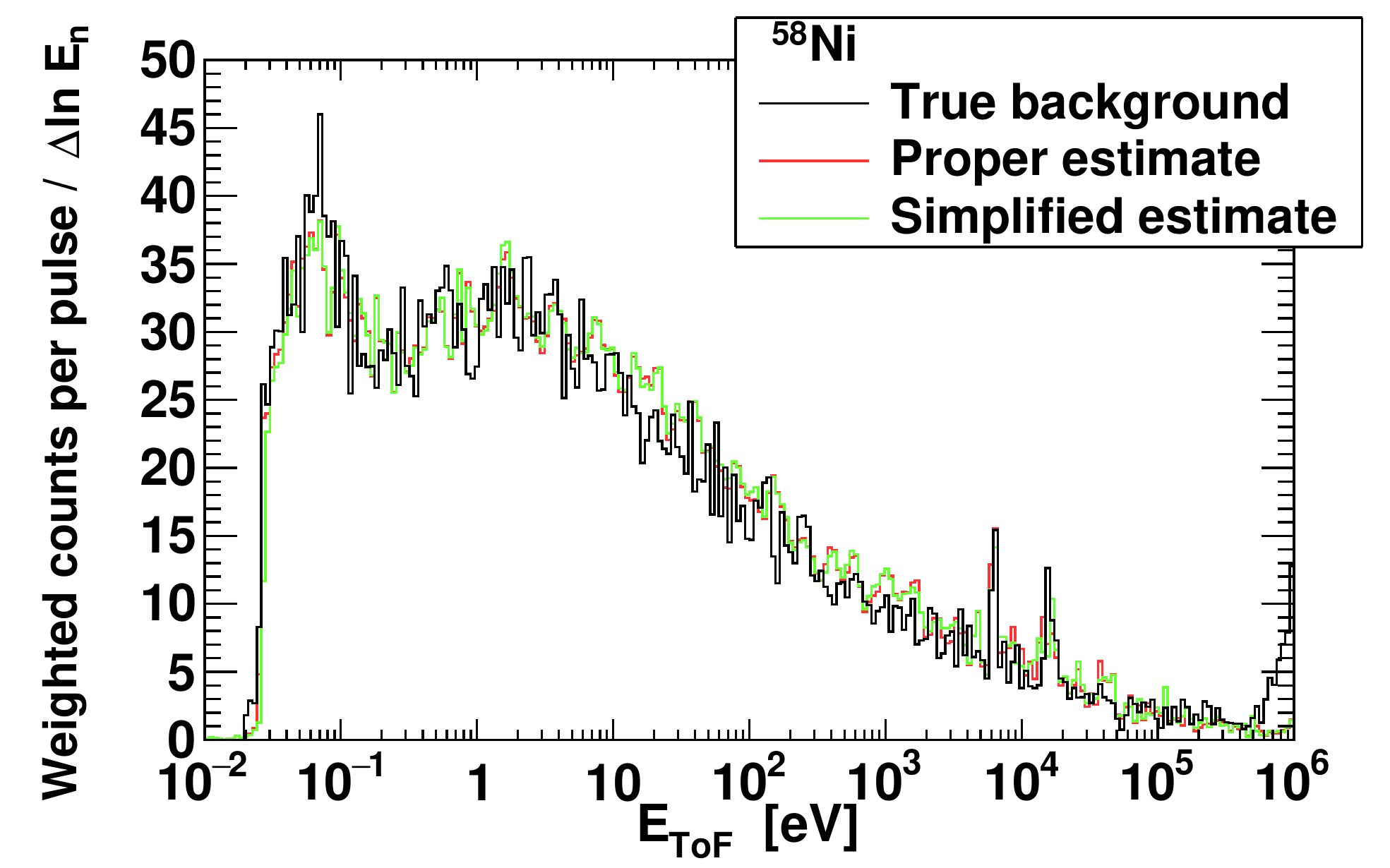}
\includegraphics[width=1.\linewidth,keepaspectratio]{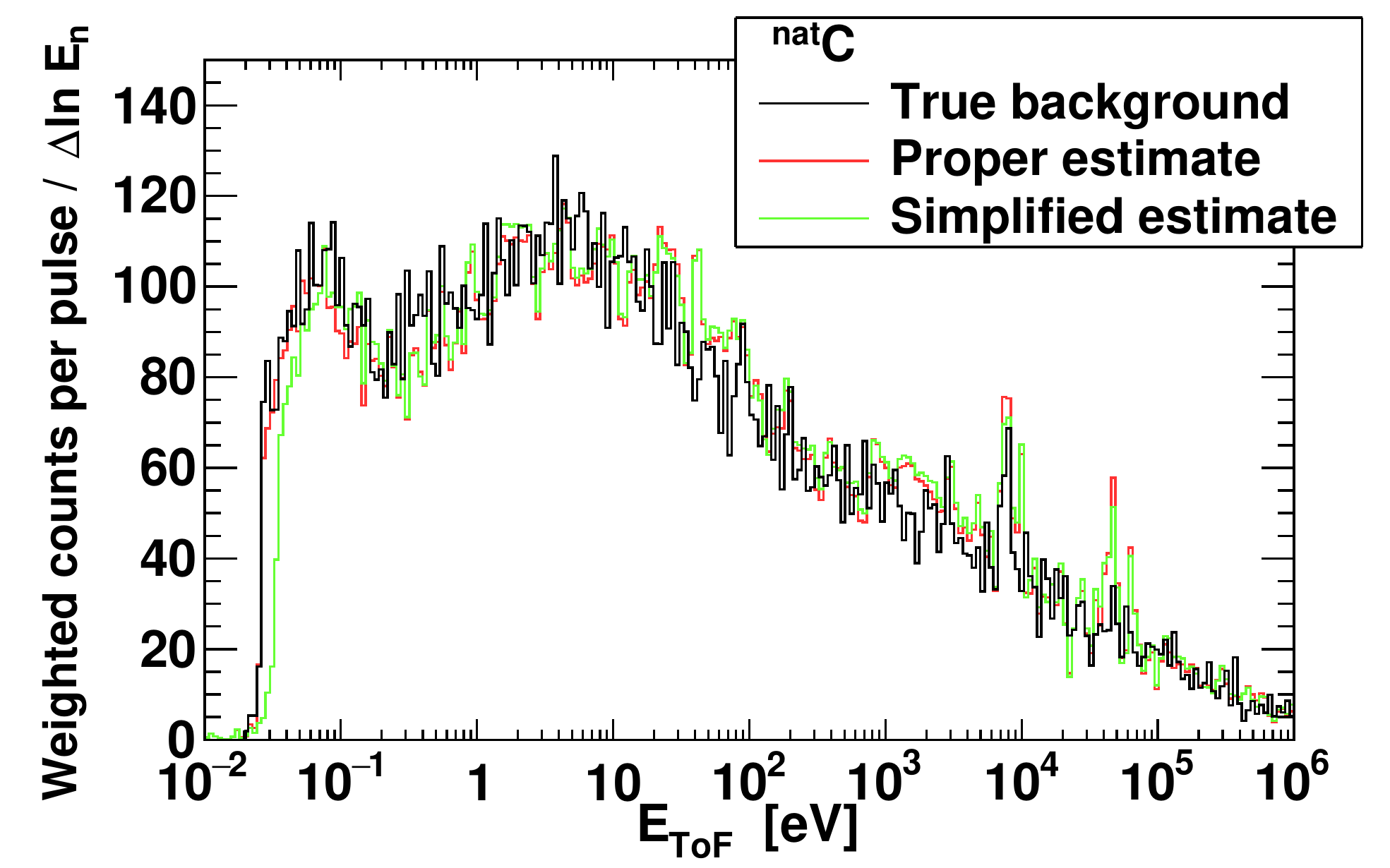}
\caption{(Color online) True neutron background of $^{58}$Ni (top panel) and $^\mathrm{nat}$C (bottom panel), compared with the background obtained by the newly proposed estimation method. The estimated background obtained by properly normalizing the simulated data, on the basis of Eq.~(\ref{weighting}), is compared against the one obtained by a simplified normalization based on Eq.~(\ref{approx}).}
\label{fig10}
\end{figure}

Since the weighting factors from Eq.~(\ref{weighting}) are dependent on more parameters than appear as the arguments of the final distribution, the weighting procedure can not be directly applied to an overall distribution of unweighted counts. Rather, it must be applied to a set of discrete data, on a count-to-count basis. \mbox{\ref{formalism}} presents the simple formalism establishing the link between the continuous distributions and the associated set of discrete data.

\begin{figure}[t!]
\includegraphics[width=1.\linewidth,keepaspectratio]{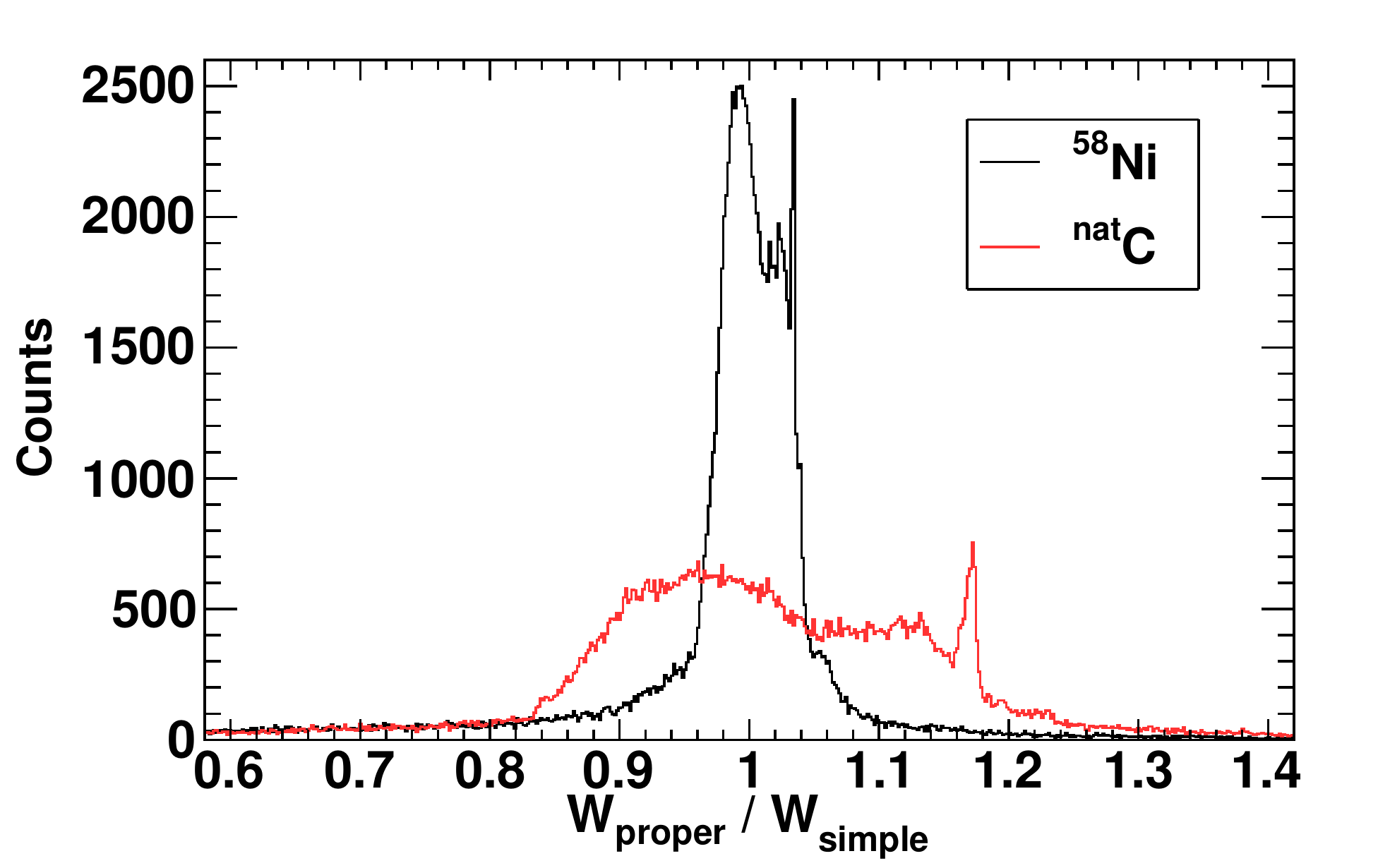}
\caption{(Color online) Distribution of ratios between the proper relativistic weighting factors $\mathcal{W}_\mathrm{proper}$ from Eq.~(\ref{weighting}) and the simplified weighting factors $\mathcal{W}_\mathrm{simple}$ from Eq.~(\ref{approx}), for $^{58}$Ni and $^\mathrm{nat}$C.}
\label{fig12}
\end{figure}

Figure~\ref{fig10} shows the neutron background for $^{58}$Ni and $^\mathrm{nat}$C, estimated using the new method, i.e. applying the weighting factors from Eq.~(\ref{weighting}). An excellent agreement with the neutron background determined from the dedicated simulations for the two samples is observed in both cases. While Eq.~(\ref{weighting}) represents the fully relativistic approach, reasonable results can also be obtained with a simplified approach in which the angular corrections and the difference in neutron energy before and after the scattering are ignored:
\begin{linenomath}\begin{equation}
\label{approx}
\mathcal{W}(*)\approx\frac{w(E_\dps)\, Y_\el(E_n)\,\phi(E_n)\,\ln\frac{E_\mx}{E_\mn}}{N_\tot}
\end{equation}\end{linenomath}
In the reconstructed-energy range \mbox{$10~\mathrm{meV}< E_\tof<1~\mathrm{MeV}$}, the ratio $\langle\mathcal{W}_\mathrm{proper}\rangle/\langle\mathcal{W}_\mathrm{simple}\rangle$ between the average proper weighting factor $\langle\mathcal{W}_\mathrm{proper}\rangle$ from Eq.~(\ref{weighting}) and the average simplified factor $\langle\mathcal{W}_\mathrm{simple}\rangle$ from Eq.~(\ref{approx}) is very close to 1 for both $^{58}$Ni and $^\mathrm{nat}$C. This may be easily understood from the fact that the major contribution to these particular backgrounds comes from essentially nonrelativistic neutrons of energies below 10~MeV, with only a minor contribution from higher energies \cite{background}. However, while the differences between the proper and the simplified weights may average out, they may become significant on the level of single counts (being more pronounced for lighter nuclei, since the boost from a neutron-nucleus center of mass frame into the laboratory frame depends on the mass of the target). This is illustrated in Fig.~\ref{fig12} for the counts from the reconstructed-energy range $10~\mathrm{meV}< E_\tof<1~\mathrm{MeV}$, for both $^{58}$Ni and $^\mathrm{nat}$C. Evidently, the corrections are mostly limited to $\pm$10\% in case of $^{58}$Ni and to $\pm$20\% in case of $^\mathrm{nat}$C, but both distributions show the long tails far beyond the central parts.

The advantage of the method proposed herein is the universality of the simulated data that need to be obtained only once for a particular experimental setup. Estimating the neutron background for a given sample requires only the proper selection of the elastic scattering cross section and the introduction of an appropriate nuclear mass into Eqs.~(\ref{primary}) and (\ref{eq4}) from \ref{considerations}, used for evaluating the central Eq.~(\ref{weighting}).

\section{Neutron background for $^{238}$U}
\label{uranium}

In this Section we present an additional complication that may affect the neutron background. The method proposed in Section~\ref{novel} considers only the elastic neutron scattering off the sample. In case of an additional background component, related to the sample properties other than the elastic scattering, there is no alternative to running the dedicated simulations specifically adapted to the particular sample. This is discussed in the example of the neutron background in a measurement of the $^{238}$U($n,\gamma$) cross section at n\_TOF \cite{intc,federica}.

\begin{figure}[t!]
\includegraphics[width=1.\linewidth,keepaspectratio]{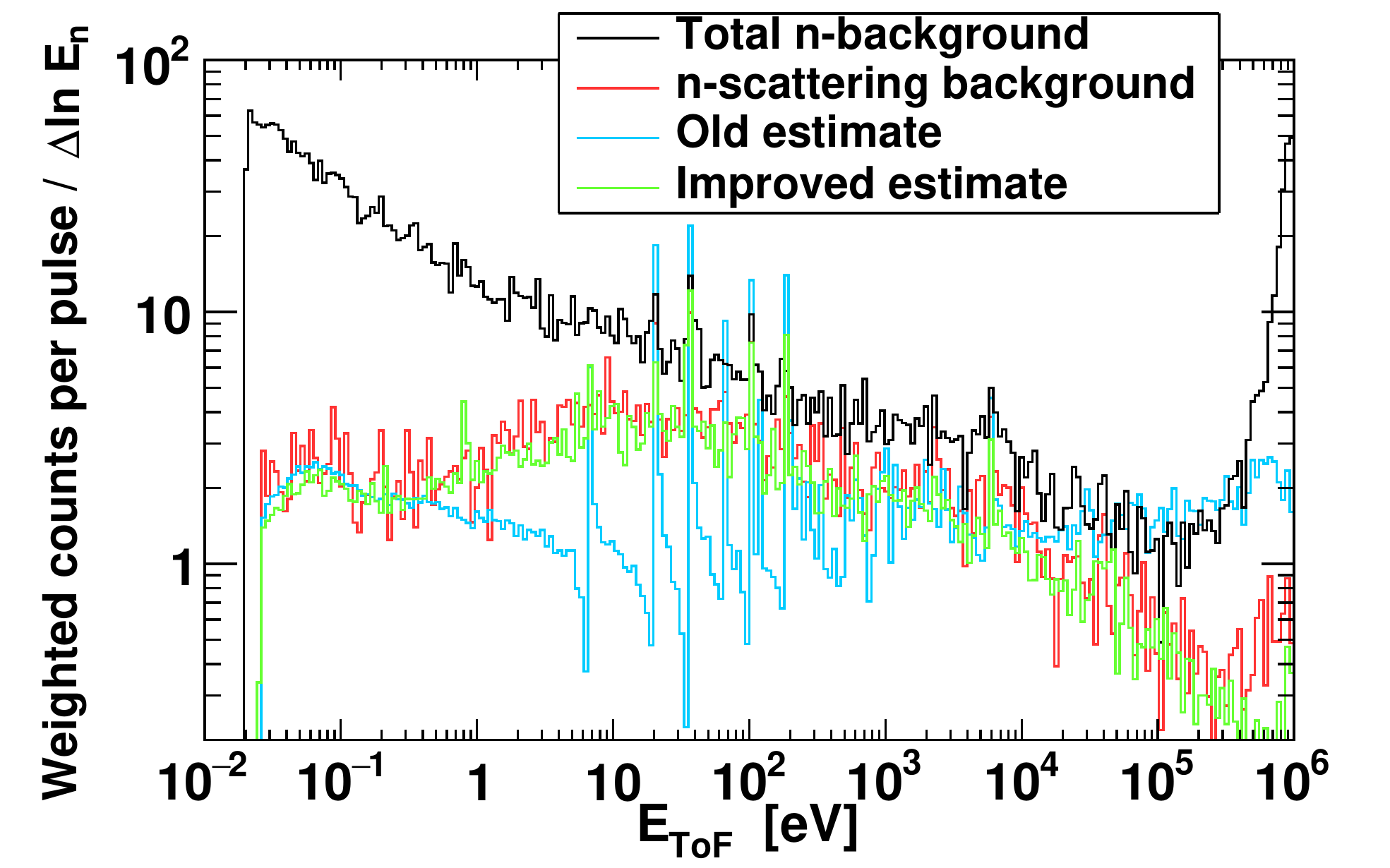}
\caption{(Color online) Total (weighted) neutron background of $^{238}$U, compared against the background caused exclusively by neutrons elastically scattered off the sample. Two background estimates are also shown: one obtained from the neutron detection efficiency of the experimental setup, the other one by the improved estimation technique (see the main text for details).}
\label{fig9}
\end{figure}

Figure~\ref{fig9} shows the simulated neutron background for a $^{238}$U sample, clearly separating a portion of the background caused exclusively by the neutron scattering off the sample. The presence of a strong additional component is immediately evident. According to the simulations, this component is caused by high energy neutrons (above 600~keV) inducing fission reactions on $^{238}$U. A variety of radioactive fission fragments is produced in the process, most so short-lived that their decay falls within $\sim$100~ms data acquisition window of the n\_TOF facility. The products of these decays (mostly $\beta$ rays) are then detected alongside the capture $\gamma$ rays, contributing to the neutron background. The detection of these secondary products is similar to the detection of $\beta$ rays from the $^{12}$C($n,p$)$^{12}$B reaction on the carbon sample. This background component, although related to the activation of the sample, can not be measured separately in beam-off runs, since the half-lives of the produced radioactive isotopes span just a few neutron bunches. The sudden increase in the total neutron background above several hundreds of keV is due to the prompt $\gamma$ rays released by neutron inelastic scattering and neutron-induced fission.

Figure~\ref{fig9} also shows the neutron background estimated by the older technique -- represented by Eq.~(\ref{estimate}) -- comparing it against the background obtained by the new technique from Section~\ref{novel}. The superiority of the new technique in reconstructing the portion of the background caused by the neutron scattering is clearly evident. Note that in the energy range above several hundred keV the angular distribution of elastically scattered neutrons from $^{238}$U departs from isotropy \cite{endf71}. While the dedicated simulations properly account for the angular distribution of the scattered neutrons, the new technique at its basic level explicitly assumes isotropic scattering in the neutron-nucleus center of mass frame, leading to a more pronounced discrepancy between the true and the estimated background. However, the correction for the realistic angular distribution may be introduced directly through the angular correction factor $\eta_\E(\cs)$. Assuming the angular distribution $\A_\E^{(\mathrm{cm})}(\cos\Theta)$ of neutrons scattered in the center of mass frame to be known and normalized to unity (with $\Theta$ as the neutron scattering angle in the neutron-nucleus center of mass frame),  the angular correction factor simply becomes:
\begin{linenomath}\begin{equation}
\label{general}
\eta_\E(\cs)=2\,\A_\E^{(\mathrm{cm})}(\cos\Theta)\times\left|\frac{\D\cos\Theta}{\D\cs}\right|
\end{equation}\end{linenomath}
thus accommodating even the most general form of the angular distribution. Of course, $\cos\Theta$ needs to be treated here as a function of $\cs$, which is achieved directly by applying Eq.~(\ref{eq4}) from \ref{considerations}.

In conclusion, the neutron background for $^{238}$U is dominated by an additional component caused by the radioactive decay of short-lived fission products (together with the fission neutrons further enhancing the neutron background). However, it should be noted that the gravity of this issue is strongly related to the length of the particular neutron flight-path ($\sim$185~m for Experimental Area 1 at n\_TOF), which determines the time-energy correlation $E_\tof\leftrightarrow T$. Thus, the only way to correctly estimate the background is by means of dedicated simulations, taking into account the full framework of the known neutron reactions induced both within and without the sample, and closely following their complete temporal evolution. Nevertheless, the newly proposed method from Section~\ref{novel} may still be used for a fast and reliable estimation of the neutron background originating exclusively from elastic neutron scattering off the sample.


\section{Summary}
\label{summary}

Following Ref. \cite{background}, we have performed a close investigation of the relationship between the neutron background in neutron capture measurements and the neutron sensitivity related to the experimental setup. The neutron background for a $^{58}$Ni sample was used in order to illustrate the difficulties in estimating the neutron background from neutron sensitivity considerations, i.e. from the neutron detection efficiency related to the experimental setup. As opposed to the neutron sensitivity being a function of the primary neutron energy, the neutron background is a function of the reconstructed neutron energy, which is affected by the temporal evolution of the neutron-induced reactions. As a consequence, the neutron background may be overestimated under the capture resonances when the estimation is attempted on a basis of the neutron sensitivity, or even on a basis of the surrogate measurements with the neutron scatterer of natural carbon. The reason is that correlations between the primary neutron energy and the reconstructed energies are neglected in considerations based only on the neutron sensitivity. Outside the resonance region a good agreement is found between the true neutron background and the background estimated from the neutron detection efficiency related to the experimental setup, and from the neutron background for natural carbon (we remind that in the Experimental Area 1 of the n\_TOF facility the pure neutron background for the carbon sample is experimentally available only above 1 keV). Therefore, these methods may still be used, provided that a smoothed elastic scattering cross section is used, instead of the resonant one.

An improved neutron background estimation technique was presented, relying on the calculation of the neutron sensitivity as a function of the reconstructed neutron energy. Supplemented by an advanced data analysis procedure, taking into account the fully relativistic kinematics of the neutron scattering, the proposed procedure yields excellent agreement between the true and the estimated neutron backgrounds for $^{58}$Ni and $^\mathrm{nat}$C.

The above considerations apply only to the background caused by elastically scattered neutrons. In the presence of reactions leading to the production of short-lived radioactive nuclides within the sample itself, an additional background component may be present. This has been illustrated in the case of the $^{238}$U sample, for which neutron-induced fission at primary neutron energies above 600~keV translates into a strong background component at reconstructed neutron energies in the thermal and epithermal region, through the detection of radioactive residuals produced by fission. The same case can also be made for the carbon sample measured in the Experimental Area 1 of the n\_TOF facility, if one considers the $^{12}$C($n,p$)$^{12}$B reaction as interfering with the neutron background measurements. In all these cases -- particularly, in measurements of capture cross sections of actinides -- the only way of properly estimating the total neutron-induced background is by means of detailed Monte Carlo simulations.\\

\textbf{ Acknowledgements}\\

This work was supported by the Croatian Science Foundation under Project No. 1680. The GEANT4 simulations have been run at the Laboratory for Advanced Computing, Faculty of Science, University of Zagreb.

\appendix

\hypertarget{response_function}{}
\section{Detector response function}
\label{response_function}

Let us suppose we have the function $f_{\E,\cs}^{(\prim)}(T)$ of the detector response to \emph{primary} neutrons of energy $\E$, scattered by an angle $\cs$ (we adopt the notation from Eq.~(\ref{short_cos}) for the cosine of the scattering angle), $f_{\E,\cs}^{(\prim)}(T)$ being a distribution of time delays $T$ between the neutron scattering off the sample (at $T=0$) and the detection of counts caused by reactions of the scattered neutrons. In parallel, let us consider the function $f_{E_n,\cs}^{(\scat)}(T)$ of the detector response to neutrons \emph{scattered} with the energy $E_n$. Evidently, the following must hold:
\begin{linenomath}\begin{equation}
\label{relation}
f_{\E,\cs}^{(\prim)}(T)\:\D\E=f_{E_n(\E,\cs),\cs}^{(\scat)}(T)\:\D E_n
\end{equation}\end{linenomath}
since the effect must be the same whether we regard the primary neutron of energy $\E$ as the ultimate source of the detected counts, or the same neutron after the scattering by the angle $\cs$, with the associated scattering energy $E_n(\E,\cs)$.

Let us now consider the contribution $\D^3N_\dtc(\E,\cs,t)$ to the detected counts caused by the primary neutrons of energy $\E$ which were scattered by the angle $\cs$, where the counts themselves are detected at time $t$ after the primary neutron production (at $t=0$). The total time delay $t$ is given by the neutron time-of-flight $\tau_\E$ along the flight-path between the neutron source (in particular, the n\_TOF spallation target) and the irradiated sample, and the time delay $T$ between the neutron scattering off the sample and detecting the count:
\begin{linenomath}\begin{equation}
t=T+\tau_\E
\end{equation}\end{linenomath}
The detector response $f_{\E,\cs}^{(\prim)}(T=t-\tau_\E)$ to the neutrons characterized by $\E$ and $\cs$ determines the number of the detected counts as:
\begin{linenomath}\begin{equation}
\label{counts}
\D^3N_\dtc(\E,\cs,t)=f_{\E,\cs}^{(\prim)}(t-\tau_\E)\, \phi(\E)\, Y_\el(\E)\, \A_\E^{(\lab)}(\cs)\, \D\E\, \D\cs \D t
\end{equation}\end{linenomath}
with $\phi(\E)$ as the neutron flux, $Y_\el(\E)$ as the yield of the sample-scattered neutrons, given by Eq.~(\ref{yield}), and $\A_\E^{(\lab)}(\cs)$ as the angular distribution of the neutrons scattered in the laboratory frame, such as the one from Eq.~(\ref{ang}). (Throughout the rest of the paper the neutron flux $\phi(\E)$ is given in units of lethargy, requiring the transition $\D\E\rightarrow\D\ln\E$.) The detector response function is extracted from the dedicated simulations. However, when the neutrons are simulated not as coming from the primary beam, but as already having been scattered off the sample -- which is a backbone of the improved method from Section~\ref{novel} -- then the simulations yield the detector response function $f_{E_n,\cs}^{(\scat)}(T)$, instead of $f_{\E,\cs}^{(\prim)}(T)$. Fortunately, the relation from Eq.~(\ref{relation}) overcomes this difficulty, allowing to rewrite Eq.~(\ref{counts}) and to express the neutron background $B(t)$ (untreated by the Pulse Height Weighting Technique) as the function of the total time delay $t$:
\begin{linenomath}\begin{align}
\begin{split}
\label{background}
&B(t)=\frac{\D N_\dtc(t)}{\D t}=\\
&=\int_{0}^{\infty}\!\!\!\D\E\int_{-1}^{1}\!\D\cs\times \frac{\partial E_n}{\partial \E}\: f_{E_n(\E,\cs),\cs}^{(\scat)}(t-\tau_\E)\, \phi(\E)\, Y_\el(\E)\, \A_\E^{(\lab)}(\cs)
\end{split}
\end{align}\end{linenomath}
Evidently, attempting to follow this procedure would be a formidable computational task, requiring the identification of the detector response $f_{E_n,\cs}^{(\scat)}(T)$ as a function of no less than three variables, with a sufficient statistical accuracy. However, the data analysis procedure laid down in Section~\ref{novel} -- relying on the proper weighting of the individual counts -- circumvents this problem by immediately building the integral from Eq.~(\ref{background}), instead of first requiring the extraction of the detailed multidimensional detector response function. The ultimate confirmation of this claim comes in form of Eq.~(\ref{ultimate}) from \mbox{\ref{formalism}}.

\hypertarget{analysis}{}
\section{Analysis of the simulated data}
\label{analysis}

\subsection{Initial considerations}
\label{considerations}

In the simulations the neutrons are treated as if already having been scattered off the sample. Though they are simulated isotropically in the laboratory frame, it is more justifiable to assume that scattering is isotropic in the center of mass frame of the primary neutron and the target nucleus. Furthermore, if the primary neutron beam was assumed to be isolethargic, the energy distribution of the scattered neutrons would not be such. Therefore, the detected counts caused by an over-idealized stream of scattered neutrons need to be weighted in a manner which will make them appear as if they were caused by an isolethargic beam of primary neutrons, which have been  scattered isotropically in the neutron-nucleus center of mass frame. After this correction, the isolethargic distribution may be reliably corrected by the actual energy dependence of the neutron flux, applying the correction to the energy distribution of the primary neutron beam, rather than to the distribution of scattered neutrons. To apply this correction properly, we will have to treat the primary neutron flux as a function of the \emph{primary energy} $\E$ that the neutron must have had in the primary beam in order to be scattered by the scattering angle $\theta$ and with the \emph{scattering energy} $E_n$. We remind that $\theta$ and $E_n$ have been independently sampled in the simulations: the scattering angle isotropically, the scattering energy isolethargically. Adopting the notation from Eq.~(\ref{short_cos}) and employing the relativistic scattering kinematics, the expression for the primary energy $\E$ may be obtained:
\begin{linenomath}\begin{align}
\label{primary}
\begin{split}
\E=&\frac{E_nc^2}{E_n\big(E_n+2mc^2\big)\cs^2-\big[E_n-(M-m)c^2\big]^2}\times\\
&\bigg\{(m+M)\big[E_n-(M-m)c^2\big]+\\
&\big(E_n+2mc^2\big)\cs\Big[\sqrt{M^2-m^2\big(1-\cs^2\big)}-m\cs\Big]\bigg\}
\end{split}
\end{align}\end{linenomath}
with $m$ as the mass of the neutron, $M$ as the mass of the scattering nucleus and $c$ as the speed of light in vacuum.

The isotropically simulated angular distribution of the scattered neutrons can be translated into the more realistic laboratory distribution (which is the relativistically transformed distribution of the neutrons isotropically scattered in the neutron-nucleus center of mass frame) by means of the angular correction factor $\eta_\E(\cs)$, as used in Eq.~(\ref{ang}):
\begin{linenomath}\begin{equation}
\label{eq3}
\eta_\E(\cs)=\left|\frac{\D\cos\Theta}{\D\cs}\right|
\end{equation}\end{linenomath}
Here $\Theta$ is the neutron scattering angle in the center of mass frame, relative to the initial beam direction. The term $\cos\Theta$ is found by employing the relativistic scattering kinematics:
\begin{linenomath}\begin{equation}
\label{eq4}
\cos\Theta=\frac{ p\cs -E\,\beta_\cm /c}{\sqrt{p^2\big(1+\beta_\cm^2\cs^2\big)+\beta_\cm^2 m^2c^2-2p E\,\beta_\cm \cs/c}}
\end{equation}\end{linenomath}
The terms required for the evaluation of Eq.~(\ref{eq4}) are given by:
\begin{linenomath}\begin{equation}
\label{eq5}
\beta_\cm=\frac{\sqrt{\E^2+2\E mc^2}}{\E+(m+M)c^2}
\end{equation}\end{linenomath}
\begin{linenomath}\begin{equation}
\label{eq6}
p=\frac{\beta_\cm c}{1-\beta_\cm^2\cs^2}\left[\frac{M \E  +m(m+M)c^2}{\E+(m+M)c^2}\cs+\sqrt{M^2-m^2\big(1-\cs^2\big)}\right]
\end{equation}\end{linenomath}
\begin{linenomath}\begin{equation}
\label{eq7}
E=c\sqrt{p^2+m^2c^2}
\end{equation}\end{linenomath}
Here $\beta_\cm$ is a conventional relativistic term for the center of mass speed in the laboratory frame (\mbox{$\beta_\cm=v_\cm/c$}, with $v_\cm$ as the actual speed). Additionally, $p$ and $E$ are the momentum and total energy of the scattered neutron in the laboratory frame.

Though by plugging Eqs.~(\ref{eq4})--(\ref{eq7}) into Eq.~(\ref{eq3}), a correction factor $\eta_\E(\cs)$ may be analytically calculated, the exact expression is long and tedious. Therefore, the derivative from Eq.~(\ref{eq3}) is best calculated numerically. Furthermore, the reader may note that the scattered neutron energy $E$ from Eq.~(\ref{eq7}) should correspond to $E=E_n+mc^2$, where the kinetic energy $E_n$ of the scattered neutron has been directly sampled in the simulations. However, the correction factor $\eta_\E(\cs)$ must be calculated for a fixed primary energy $\E$, instead of $E_n$. Hence, during the calculation of the correction factor $\eta_\E(\cs)$, the momentum $p$ from Eq.~(\ref{eq6}) and energy $E$ from Eq.~(\ref{eq7}) must be treated as functions of $\E$ and $\cs$, and have to be varied accordingly .

\subsection{Derivation of the weighting factors}
\label{derivation}

For brevity of expressions, we adopt the notation $*$ from Eq.~(\ref{star}) for the set of all relevant parameters. The first step in determining the weighting factors $\mathcal{W}(*)$, dependent on any combination of these parameters, is the normalization of the data by the total number $\delta^2N_\sml(E_n,\cs)$ of neutrons simulated with scattering energy $E_n$ and scattering angle $\cs\!$. To obtain the contribution to the detected counts per single neutron bunch, the statistical effect of a single scattered neutron -- isolated by the previous normalization -- must be amplified by the number $\delta^2N_\prim(E_n,\cs)$ of primary neutrons (those from the primary beam) that can be scattered by the angle $\cs\!$, with the energy $E_n$. Furthermore, it is necessary to account for the probability that the primary neutron of energy $\E$ will indeed be scattered. This probability may be expressed as the yield $Y_\el(\E)$ of elastically scattered neutrons from Eq.~(\ref{yield}). Finally, since we are applying the Pulse Height Weighting Technique, the most evident weighting factor is given by the weighting function $w(E_\dps)$, dependent on the energy $E_\dps$ deposited in the detector. Combined, the previous considerations give rise to the total weighting factor $\mathcal{W}(*)$ that has to be applied to each detected count:
\begin{linenomath}\begin{equation}
\label{w_start}
\mathcal{W}(*)=w(E_\dps)\, Y_\el\big(\E)\times\frac{\delta^2N_\prim(E_n,\cs)}{\delta^2N_\sml(E_n,\cs)}
\end{equation}\end{linenomath}
Since the scattered neutrons have been simulated isolethargically and isotropically (in the laboratory frame), the term $\delta^2N_\sml(E_n,\cs)$ is simply determined as: 
\begin{linenomath}\begin{equation}
\label{term1}
\delta^2N_\sml(E_n,\cs)=\frac{N_\tot}{\ln\frac{E_\mx}{E_\mn}}\delta\ln E_n\times\frac{\D\cs}{2}
\end{equation}\end{linenomath}
where $N_\tot$ is the total number of simulated neutrons, with $E_\mn$ and $E_\mx$ as the minimum and maximum sampling energies, respectively. The angular distribution of neutrons scattered from the primary beam -- under the assumption of isotropic scattering in the neutron-nucleus center of mass frame -- is already known from Eq.~(\ref{ang}). With the neutron neutron flux $\phi(\E)$ conveniently given in units of lethargy, the term $\delta^2N_\prim(E_n,\cs)$ is also easily expressed:
\begin{linenomath}\begin{equation}
\label{term2}
\delta^2N_\prim(E_n,\cs)=\phi(\E)\delta\ln \E\times\eta_\E(\cs)\frac{\D\cs}{2}
\end{equation}\end{linenomath}
Combined, Eqs.~(\ref{term1}) and (\ref{term2}) lead to:
\begin{linenomath}\begin{equation}
\label{ratio}
\frac{\delta^2N_\prim(E_n,\cs)}{\delta^2N_\sml(E_n,\cs)}=\frac{\ln\frac{E_\mx}{E_\mn}}{N_\tot}\,\phi(\E)\,\eta_\E(\cs)\times\frac{\delta\ln\E}{\delta \ln E_n}
\end{equation}\end{linenomath}
Since the primary neutron energy $\E$ may be calculated for any combination of the scattering energy and angle (up to the possible kinematic limitations, dependent on the projectile and target masses),  $E_n$ and $\cs$ are independent, making the derivative from Eq.~(\ref{ratio}) a partial one:
\begin{linenomath}\begin{equation}
\label{der}
\frac{\delta\ln\E}{\delta \ln E_n}=\frac{E_n}{\E}\frac{\partial\E}{\partial E_n}
\end{equation}\end{linenomath}
Finally, plugging Eqs.~(\ref{ratio}) and (\ref{der}) back into Eq.~(\ref{w_start}) yields the full expression for the overall weighting factor:
\begin{linenomath}\begin{equation}
\mathcal{W}(*)=\frac{w(E_\dps)\, Y_\el(\E)\,\phi(\E)\,\eta_\E(\cs)\,\ln\frac{E_\mx}{E_\mn}}{N_\tot}\times\frac{E_n}{\E}\frac{\partial\E}{\partial E_n}
\end{equation}\end{linenomath}
We remind that the derivative $\partial\E/\partial E_n$ may be easily calculated from Eq.~(\ref{primary}).

\hypertarget{formalism}{}
\section{Simple formalism linking the discrete data and continuous distributions}
\label{formalism}

We present a convenient formalism establishing the link between discrete data and continuous distributions that are built from these data. This formalism is especially useful when each datum must be applied its own weighting factors, dependent on more parameters or other parameters than those that appear as the arguments of the final distribution. We demonstrate the formalism by immediately applying it to the estimated neutron background from Section~\ref{novel}. For this purpose, let us consider the contribution $\delta^4N_\dtc^{(\mathcal{W})}(*)$ to the detected counts weighted by the weighting factors $\mathcal{W}(*)$, dependent on any combination of the physical parameters. As before, we adopt the notation from Eqs.~(\ref{short_cos}) and (\ref{star}). Within the formalism of continuous distributions, the contribution $\delta^4N_\dtc^{(\mathcal{W})}(*)$ to the detected counts from an element of the parameter space defined by $\delta E_\tof$, $\delta E_n$, $\delta E_\dps$ and $\delta \cs$, may be written as:
\begin{linenomath}\begin{align}
\label{rep1}
\begin{split}
\delta^4N_\dtc^{(\mathcal{W})}(*)\sim\frac{\D^4N_\dtc^{(\mathcal{W})}(E_\tof;\:E_n,E_\dps,\cs)}{\D E_\tof\times \D E_n\times\D E_\dps\times \D\cs}\times&\\
\delta E_\tof\times \delta E_n&\times\delta E_\dps\times \delta\cs
\end{split}
\end{align}\end{linenomath}
By $\sim$ we have denoted the relation of \emph{correspondence}, i.e. the \emph{representation} of the term from one side by the term from the other side. In that, it holds:
\begin{linenomath}\begin{equation}
\D^4N_\dtc^{(\mathcal{W})}(*)=\mathcal{W}(*)\times\D^4N_\dtc(*)
\end{equation}\end{linenomath}
with $\D^4N_\dtc(*)$ as the elementary contribution to the unweighted counts. On the other hand, when building the distribution from discrete data, the contribution $\delta^4N_\dtc^{(\mathcal{W})}(*)$ appears as a union of all appropriately weighted counts $\mathcal{C}_i\big(E_\tof;\:\mathcal{W}(*)\big)$ from an associated element of the parameter space:
\begin{linenomath}\begin{equation}
\label{rep2}
\delta^4N_\dtc^{(\mathcal{W})}(*)\sim\bigcup_{i=1}^{\delta^4 N_\dtc(*)}\mathcal{C}_i\big(E_\tof;\:\mathcal{W}(*)\big)
\end{equation}\end{linenomath}
with $\delta^4 N_\dtc(*)$ as the total number of detected counts from the parameter space volume $\delta E_\tof\times \delta E_n\times\delta E_\dps\times \delta\cs$. In Eqs.~(\ref{rep1}) and (\ref{rep2}) $E_\tof$ holds a prominent place because we are expressing the final distribution (the neutron background from Section~\ref{novel}) as a function of the reconstructed energy. Combining Eqs.~(\ref{rep1}) and (\ref{rep2}) leads to a direct correspondence between the set of discrete data and their representation by the continuous distribution:
\begin{linenomath}\begin{align}
\label{master}
\begin{split}
\frac{\D^4N_\dtc^{(\mathcal{W})}(E_\tof;\:E_n,E_\dps,\cs)}{\D E_\tof\times \D E_n\times\D E_\dps\times \D\cs}\times\delta E_\tof\times \delta E_n\times\delta E_\dps\times \delta\cs\sim&\\
\bigcup_{i=1}^{\delta^4 N_\dtc(*)}\mathcal{C}_i\big(E_\tof;\:\mathcal{W}(*)\big)&
\end{split}
\end{align}\end{linenomath}
In this respect, the integration operator from the continuous distribution domain also has a clearly defined union-of-counts counterpart within the discrete data domain:
\begin{linenomath}\begin{equation}
\int_{x_1}^{x_2}\D x\sim\bigcup_{x=x_1}^{x_2}
\end{equation}\end{linenomath}
This enables an explicit insight into the structure of the final weighted distribution we are expecting to build:
\begin{linenomath}\begin{equation}
\label{ultimate}
\frac{\D N_\dtc^{(\mathcal{W})}(E_\tof)}{\D\ln E_\tof}\sim\frac{1}{\delta\ln E_\tof}
\bigcup_{E_n}\bigcup_{E_\dps}\bigcup_{\cs}\bigcup_{i=1}^{\delta^4 N_\dtc(*)}\mathcal{C}_i\big(E_\tof;\:\mathcal{W}(*)\big)
\end{equation}\end{linenomath}
Here we have immediately made a transition $\D E_\tof\rightarrow \D\ln E_\tof$ simply because in Section~\ref{novel} we have used the histogram binning uniformly distributed over the logarithmic scale.




\begin{thebibliography}{00}
\bibitem{koehler} P. E. Koehler, R. R. Winters, K. H. Guber, et al., Phys. Rev. C 62 (2000) 055803.
\bibitem{plag} R. Plag, M. Heil, F. K\"{a}ppeler, et. al., Nucl. Instr. and Meth. A 496 (2003) 425.
\bibitem{carlos} C. Guerrero, A. Tsinganis, E. Berthoumieux, et al., Eur. Phys. J. A 49 (2013) 27.
\bibitem{ear2_1} C. Wei{\ss}, E. Chiaveri, S. Girod, et al., Nucl. Instr. and Meth. A 799 (2015) 90.
\bibitem{ear2_2} S. Barros, I. Bergstr\"{o}m, V. Vlachoudis and C. Wei\ss, J. Instrum. 10 (2015) P09003.
\bibitem{geant4} S. Agostinelli, J. Allison, K. Amako, et al., Nucl. Instr. and Meth. A 506 (2003) 250.
\bibitem{background} P. \v{Z}ugec, N. Colonna, D. Bosnar, et al., Nucl. Instr. and Meth. A 760 (2014) 57.
\bibitem{ni58} P. \v{Z}ugec, M. Barbagallo, N. Colonna, et al., Phys. Rev. C 89 (2014) 014605.
\bibitem{carbon} P. \v{Z}ugec, N. Colonna, D. Bosnar, et al., Phys. Rev. C 90 (2014) 021601(R).
\bibitem{guber} K. H. Guber, H. Derrien, L. C. Leal, et al., Phys. Rev. C 82 (2010) 057601.
\bibitem{frank} F. Gunsing, E. Berthoumieux, G. Aerts, et al., Phys. Rev. C 85 (2012) 064601.
\bibitem{pwht} R. L. Macklin, J. H. Gibbons, Phys. Rev. 159 (1967) 1007.
\bibitem{pwht_ntof} U. Abbondanno, G. Aerts, H. Alvarez, et al., Nucl. Instr. and Meth. A 521 (2004) 454.
\bibitem{endf71} M. B. Chadwick, M. Herman, P. Oblo\v{z}insk\'{y}, et al., Nucl. Data Sheets 112 (2011) 2887.
\bibitem{intc} The n\_TOF Collaboration, \emph{Neutron capture cross section measurements of $^{238}$U, $^{241}$Am and $^{243}$Am at n\_TOF}, Proposal to the ISOLDE and Neutron Time-of-Flight Committee, CERN-INTC-2009-025 / INTC-P-269 20/04/2009.
\bibitem{federica} F. Mingrone, C. Massimi, S. Altstadt, et al., Nucl. Data Sheets 119 (2014) 18.
\end{thebibliography}
\end{document}